\documentclass[twocolumn,aps,raggedbottom,nobalancelastpage,amssymb,floatfix]{revtex4-1}

\usepackage{graphicx}
\usepackage{amsmath}
\usepackage{amssymb}
\usepackage{bm}
\usepackage{pgf}
\usepackage{color}
\usepackage{amsfonts}
\usepackage{hyperref}

\begin{document}

\title{Learning phase transitions by confusion}
\author{Evert P.L. van Nieuwenburg*}
\affiliation{Institute for Theoretical Physics, ETH Zurich, 8093 Z{\"u}rich, Switzerland}
\author{Ye-Hua Liu}
\affiliation{Institute for Theoretical Physics, ETH Zurich, 8093 Z{\"u}rich, Switzerland}
\author{Sebastian D. Huber}
\affiliation{Institute for Theoretical Physics, ETH Zurich, 8093 Z{\"u}rich, Switzerland}

\begin{abstract}
  Classifying phases of matter is a central problem in physics. For quantum mechanical systems, this task can be daunting owing to the exponentially large Hilbert space. Thanks to the available computing power and access to ever larger data sets, classification problems are now routinely solved using machine learning techniques. Here, we propose to use a neural network based approach to find phase transitions depending on the performance of the neural network after training it with deliberately incorrectly labelled data. We demonstrate the success of this method on the topological phase transition in the Kitaev chain, the thermal phase transition in the classical Ising model, and the many-body-localization transition in a disordered quantum spin chain. Our method does not depend on order parameters, knowledge of the topological content of the phases, or any other specifics of the transition at hand. It therefore paves the way to a generic tool to identify unexplored phase transitions.
\end{abstract}

\maketitle


\section{Introduction}
Machine learning as a tool for analyzing data is becoming more and more prevalent in an increasing number of fields~\cite{Jordan2015}. This is due to a combination of availability of large amounts of data, and the advances in hardware and computational power (most notably through the use of GPUs).

Two typical methods of machine learning can be distinguished, namely the unsupervised and supervised methods. In the former, the algorithm receives no input other than the data and is asked, e.g., to extract features or to cluster the samples. Such an unsupervised approach was applied to the physical problem of identifying phase transitions and order parameters, from images of classical configurations of Ising models \cite{Wan16a}. In the supervised learning methods, the data has to be supplemented by a labelling. Typical problems in this direction involve the classification of many samples, where each sample is assigned a class-label. The machine is trained to recognize samples and predict their associated label, demonstrating that it has \emph{learned} by doing so via correctly predicting samples it has not encountered before. This approach, too, has been demonstrated on Ising models \cite{Car16a}. Such approaches have also recently provided promising prospects for strongly correlated fermions~\cite{Chng2016, Li2016b} and the fermion sign problem~\cite{Broecker2016}. Last, we mention a slightly orthogonal approach based on reinforcement learning, where the wavefunction was represented as a particular type of artificial neural network \cite{Carleo16a}. A variational approach then allows for finding groundstates and performing time-evolution, and outperformed other state-of-the-art numerical methods for the two-dimensional Heisenberg spin model. Some topological states have been shown to have efficient and exact representations as an artificial neural network~\cite{Deng2016}.

Conversely, concepts from physics have also found their way into the field of machine learning. Examples of this are e.g. the relations between neural networks and statistical Ising models and renormalisation flow \cite{Pan14a}, the use of tensor network techniques to train them \cite{Sto16a}, and indeed the very concept of phase transitions themselves \cite{Saitta2010}.

Motivated by previous studies, we apply machine-learning techniques to the detection of phase transitions. In contrast to the previous works, however, we focus on a combination of supervised and unsupervised techniques. In most cases namely, it is exactly the labelling that one would like to find out (i.e. classification of phases). That implies that a labelling is not known beforehand, and hence supervised techniques are not directly applicable. In this Letter we demonstrate that it is possible to find the correct labels, by purposefully mislabelling the data and evaluating the performance of the machine learner. We will base our method on neural networks, which are capable of fitting arbitrary non-linear functions \cite{Hay98a}. Indeed, if a linear feature extraction method worked, there would have been no need to explicitly find labels in the first place. 

\begin{figure*}[t]
  \begin{center}
    \includegraphics[scale=0.75]{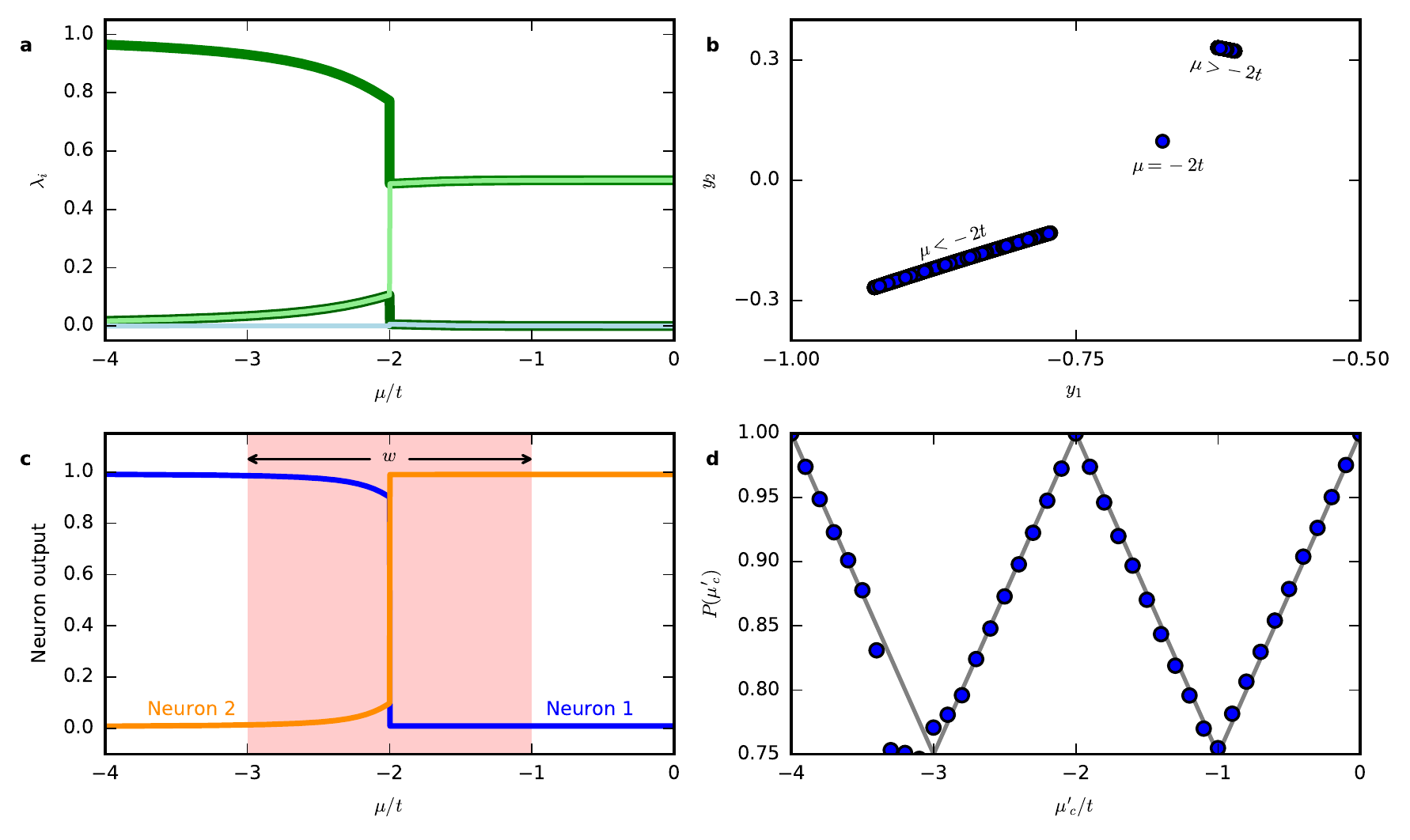}
    \caption{(\textbf{a}) Evolution of the entanglement spectrum of the Kitaev chain as a function of the chemical potential $\mu$. Here we plot the largest four eigenvalues of the reduced density matrix $\rho_A$. The degeneracy structure is clearly observable. (\textbf{b}) Principal component analysis of the entanglement spectrum. All data points are shown in the plane of the first two principal components $y_1$ and $y_2$. (\textbf{c}) Supervised learning with blanking. The shaded region is blanked out during the training phase, and the neural network can still predict the correct transition point $\mu=-2t$. (\textbf{d}) $P(\mu'_{c})$, evolution of the accuracy of prediction, as a function of the proposed critical point $\mu'_{c}$, which shows the universal W-shape. See text for more details. (Parameters for training: batch size $N_b=100$, learning rate $\alpha=0.075$ and regularization $l_2=0.001$.)
      \label{fig:kitaev}}
  \end{center}
\end{figure*}

For quantum phase transitions, one tries to learn the quantum-mechanical wavefunction $|\psi\rangle$, which contains exponentially many coefficients with increasing system size. As has been noted before \cite{Car16a}, a similar problem exists in the field of machine learning: the number of samples in a dataset has to increase exponentially with the number of features one is trying to extract. To prevent having to deal with exponentially large wavefunctions, we pre-process the data in the form of the entanglement spectrum (ES) \cite{Li2008}, which has been shown to contain important information about $|\psi\rangle$ \cite{Laflorencie2015}, although care has to be taken when interpreting phase transitions \cite{Cha14a}. 

To justify the use of the ES, we note that recently the quantum entanglement has taken up a major role in the characterization of many-body quantum systems \cite{Amico2008,Laflorencie2015}. In particular, the ES has been used as an important tool in, for example, fingerprinting topological order \cite{Thomale2010,Qi2012,Pol11a}, tensor network properties \cite{Cirac2011,Schuch2013}, quantum critical points, symmetry breaking phases \cite{Calabrese2008a,Alba2012a}, and even many-body localization \cite{Yan15a,Geraedts2016}. Very recently, an experimental protocol for measuring the ES has been proposed \cite{Pic16a}. On the level of the ES, the information of phases is not clearly identifiable as in the classical images, which we will show in the following sections. However, patterns in the ES suggest that learning and generalization is still possible. 

In the following, we will first consider the Kitaev chain as a demonstration of our method. The Kitaev chain serves as an excellent example since analytical results are available, and the ES shows a clear distinction between the two phases of the model~\cite{Kitaev2001}. We demonstrate the generalizing power of the neural network by blanking out the training data around the transition, and show that it can still predict the transition accurately. We then purposefully mislabel the data, thereby confusing the network, and introduce the characteristic shape of the networks' performance function. Next, this confusion method is demonstrated on a classical two-dimensional Ising model to show that it is not specific to the Kitaev model. Last, we apply the proposed method to the nontrivial task of finding the many-body-localization (MBL) transition from many disordered samples of a one-dimensional spin chain. We finish with a discussion and outlook.

\section{Results}
We demonstrate the various machine-learning methods on the model of the Kitaev chain:
\begin{align}
  \hat{H} = -t\sum_{i=1}^L \left(  \hat{c}_{i+1}^{\dagger}\hat{c}_{i} 
  + \hat{c}_{i+1}\hat{c}_{i} + h.c. \right)
  - \mu\sum_{i=1}^L \hat{c}_{i}^{\dagger}\hat{c}_{i}, \label{eq:KitaevH}
\end{align}
where $t>0$ controls the hopping and the pairing of spinless fermions alike and $\mu$ is a chemical potential. The groundstate of this model has a quantum phase transition from a topologically trivial ($|\mu|>2t$) to a nontrivial state ($|\mu|<2t$) as the chemical potential $\mu$ is tuned across $\mu = \pm2t$.

We use the ES to \emph{compress} the quantum mechanical wavefunction. The ES is defined as follows. The whole system is first divided into two subsets $A$ and $B$, after which the reduced density matrix of subset $A$ is calculated by partially tracing out the degrees of freedom in $B$, i.e. $\rho_A = \text{Tr}_B |\psi\rangle\langle\psi|$. Denoting the eigenvalues of $\rho_A$ as $\lambda_i$, the entanglement spectrum is then defined as the set of numbers $-\ln \lambda_i$. It is important to remark that various types of bipartition of the whole system into subsets $A$ and $B$ exist, such as dividing the bulk in extensive disconnected parts \cite{Hsieh14}, divisions in momentum space \cite{Thomale2010a} or indeed even random partitioning \cite{Vijay15}. In this work, we use the usual spatial bipartition into left and right halves of the whole system. 

As shown in Fig.~\ref{fig:kitaev}a, the entanglement spectrum of the Kitaev chain is clearly distinguishable in the two phases, especially since the nontrivial phase has a degeneracy structure as do all symmetry protected topological phases \cite{Pol11a}. This feature is clear also for human eyes, and a machine-learning routine seems to be an overkill. We use this model for demonstration purposes and in the following, we will apply the introduced methodology to more complex models. The data for machine learning is chosen to be the largest 10 eigenvalues $\lambda_i$, for $L = 20$ with an equal partitioning $L_A=L_B=10$, and for various values of $-4t\le\mu\le0$.

\subsection{Unsupervised learning}
First we perform unsupervised learning, using an established method for feature extraction. The entanglement spectra are interpreted as points in a $10$-dimensional space, and we use principal component analysis (PCA) \cite{Pearson1901} to extract mutually orthogonal axes along which most of the variance of the data can be observed. PCA amounts to a linear transformation $Y = XW$, where $X$ is an $N \times 10$ matrix containing the entanglement spectra as rows ($N = 10^4$ is the number of samples). 

The orthogonal matrix $W$ has vectors representing the principal components $\omega_\ell$ as its columns, which are determined through the eigenvalue equation $X^T X \omega_\ell = \lambda_\ell \omega_\ell$. The eigenvalues $\lambda_\ell$ are the singular values of the matrix $X$, and are hence non-negative real numbers, and we normalize them s.t. $\sum \lambda_\ell = 1$. The result of PCA is shown in Fig.~\ref{fig:kitaev}(b), and it is indeed possible to cluster the spectra in to three sets: $\mu < -2t$, $\mu = -2t$, and $\mu > -2t$.

\subsection{Supervised learning}
Although it is as a whole unnecessary, since the PCA analysis already manages to extract the phases and the transition point, we still train a feedforward neural network (NN) on the $10$-dimensional inputs. This is for demonstration only, since we have in mind the application to models in which PCA is insufficient.

We train the network with 80 hidden sigmoid neurons in a single hidden layer, and 2 output neurons. The first/second output neuron predicts the (not necessarily normalized) probability for the data to be in trivial/nontrivial phase, and the predicted phase is the phase with the larger probability. We use stochastic gradient descent and L2 regularization to try and minimize a cross-entropy cost function (for details, see e.g. Ref.~\cite{Nielsen2015}). The network easily learns to distinguish the spectra when trained on a subset of the data points and asked to predict the others.

Arguably the most important objective of machine-learning in general is that of \emph{generalization}. After all, learning is demonstrated by being able to perform well on examples that have not been encountered before. By having trained the network only on a subset of the data, and having it correctly predict others, it has already demonstrated learning. 

As another display of the generalizing power of the network, we \emph{blank out} the data in a width $w$ around $\mu = -2t$ and ask the network to interpolate and find the transition point. Figure~\ref{fig:kitaev}c shows that the network has no difficulties doing so even for $w = 2t$. We were able to go up to widths $w=3t$ before training became unreliable.

\subsection{Confusion scheme}
The PCA as an unsupervised learning technique may be applied without perfectly known 
information of the system, but it is a linear analysis and is hence incapable of extracting 
non-linear relationships among the data. On the other hand, a NN is capably of fitting any 
non-linear function \cite{Hay98a}, but a training phase with \emph{correctly} labelled 
input-output pairs is needed. In the following, we propose a scheme combining the two 
methods which we refer to as a \emph{confusion scheme}. This scheme is the main result of 
this work. 

Suppose all data lies in the parameter range $(a,b)$, and we know there exists a critical 
point $a<c<b$ such that the data could be classified into two groups. However, we do not 
know the value of $c$. We propose a critical point $c'$, and train a network that we call 
$\mathcal{N}_{c'}$ by labelling all data with parameters smaller than $c'$ with label $0$ 
and the others with label $1$. Next, we evaluate the performance of $\mathcal{N}_{c'}$ on 
the entire data set and refer to its total performance, with respect to the proposed 
critical point $c'$, as $P(c')$. We will show that the function $P(c')$ has a universal 
\emph{W-shape}, with the middle peak at the correct critical point $c$. Applying this to 
the Kitaev model, we can see from Fig.~\ref{fig:kitaev}d that for $-4t<\mu<0$, the 
prediction performance from confusion scheme has a W-shape with the middle peak at 
$\mu=-2t$. 

The W-shape can be understood as follows. We assume that the data has two different 
structures in the regimes below $c$ and above $c$, and that the NN is able to find and 
distinguish them. We refer to these different structures as features. When we set $c'=a$, 
the NN chooses to assign label $1$ to both features and thus correctly predicts $100$\% of 
the data. A similar analysis applies to $c'=b$, except that every data point is assigned 
the label $0$. When $c'=c$ is the correct labelling, the NN will choose to assign the right 
label to both sides of the critical point and again performs perfectly. When $a<c'<c$, in 
the training phase the NN sees data with the \emph{same} feature in the ranges from $a$ to 
$c'$ and from $c'$ to $c$, but having \emph{different} labels (hence the confusion). In 
this case it will choose to learn the label of the majority data, and the performance will 
be
\begin{align}
  P(c')=1-\frac{\min\left(c-c',c'-a\right)}{c-a}.
\end{align}
Similar analysis applies to $c<c'<b$. This gives the typical W-shape seen in Fig.~\ref{fig:kitaev}d. Notice that if the point $c$ is not exactly centered between $a$ and $b$, the W-shape will be slightly distorted. Its middle peak always corresponds to the correct labelling, but the depth of the minima will differ between the left and right.

\begin{figure}[t]
  \begin{center}
    \includegraphics[width=80mm]{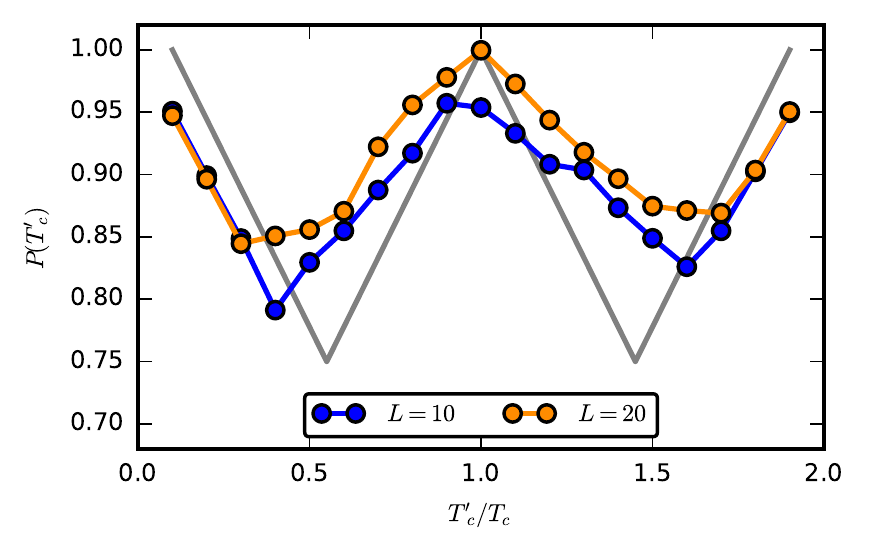}
    \caption{Position of the middle peak in the universal W-shape deviates from $T'_{c}=T_c$ for $L=10$ due to the finite-size effect. Here $k_B T_c \approx 2.27J$ is the exact transition temperature in the thermodynamic limit. For $L=20$ the middle peak is located exactly at $T'_{c}=T_c$. (Parameters for training: batch size $N_b=100$, learning rate $\alpha=0.02$ and regularization $l_2=0.005$.)
      \label{fig:Ising}} 
  \end{center}
\end{figure}

We test the confusion scheme on the thermal phase transition in the two-dimensional classical Ising model~\cite{Onsager1944}, which has been studied by both supervised learning \cite{Car16a} and unsupervised learning \cite{Wan16a} methods. Here we train a NN (with $L^2$ neurons in the input and hidden layers, and 2 neurons in the output layer) on the $L \times L$ classical configurations sampled from Monte Carlo simulation. As shown in Fig.~\ref{fig:Ising}, the W-shape again predicts the right transition temperature. Note the confusion scheme works better when the underlying feature in the data is shaper, i.e. for the larger system size $L=20$.

To confirm that the confusion scheme indeed extracts nontrivial features from the input data. We have checked the performance curve from confusion scheme, when the NN is trained on unstructured random data. We use a fictive parameter as a tuning parameter, but have completely unstructured (random) data as a function of it. Hence, the network will not find structure in the data, and a correct labelling does not exist. The middle peak of the characteristic W-shape disappears, turning it into a V-shape.

We notice that the choice of the learning rate ($\alpha$) and regularization ($l_2$) is essential for a successful training. The use of regularization is expected to reduce overfitting and make the network less sensitive to small variations of the data, hence forcing it to rather learn its structure~\cite{Nielsen2015}. However, the confusion scheme depends solely on the ability of finding the majority label for the underlying structure in the data. In this sense, overfitting is not necessarily bad. Indeed we have observed that training with a negative $l_2$ may lead to an equally good performance. We speculate that this is because a negative $l_2$ tries to quickly increase the weights, making it harder for the network to change its opinion about data samples in later stages. If the initial training data is uniformly sampled, meaning the majority data is indeed represented by a majority, the network will rapidly adjust its weights to this majority. Lastly, we mention that trainings are performed in epochs. In each epoch all training data is passed once in batches of size $N_b$ in a random order. The training is stopped when a clear W-shape is formed.

\begin{figure}[t]
  \begin{center}
    \includegraphics[width=80mm]{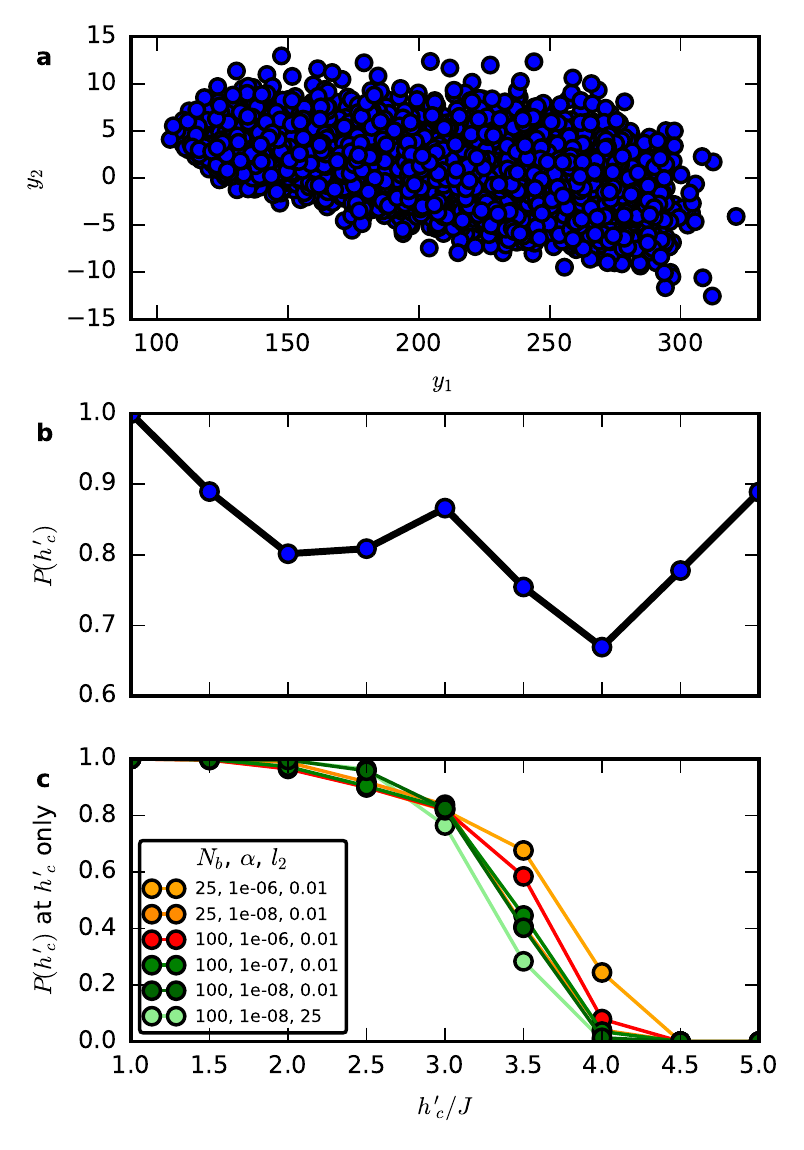}
    \caption{(\textbf{a}) Principal component analysis of the random-field Heisenberg model. Unlike in the Kitaev model or for the Ising data \cite{Wan16a}, there is no clearly observable clustering. (\textbf{b}) The characteristic W-shape of the performance curve on the MBL data. The result shows that the network $\mathcal{N}_{h'_{c}}$ for $h'_{c} \approx 3J$ performs best, indicating that this is the correct labelling. The distinction between the thermalizing and non-thermalizing phase can hence be put at $h_c \approx 3J$, in agreement with Ref.~\cite{Geraedts2016}. (Parameters for training: batch size $N_b = 100$, learning rate $\alpha = 10^{-8}$ and regularization $l_2 = 0.01$.) (\textbf{c}) The performance of network $\mathcal{N}_{h'_{c}}$, when evaluated at the point $h'_c$ only instead of on the full data, for various different sets of learning parameters (see legend). Clearly the performance of the network is most independent of the exact training scheme at $h'_c \approx 3J$, showing a robustness of this correct labelling against variations in training.
      \label{fig:MBLPerformance} \label{fig:MBLPCA}}
  \end{center}
\end{figure}

\subsection{Random-field Heisenberg chain}
We will now test our proposed scheme on an example where we the exact location of the transition point is now known~\cite{Nandkishore2015}. We study a case of interest in recent literature, namely that of many-body localization. We consider the following model:
\begin{align}
  H = J \sum_{i=1}^{L} \boldsymbol{S}_i\cdot\boldsymbol{S}_{i+1} + \sum_{\alpha=x,y,z}\sum_{i=1}^L h^\alpha_i S^\alpha_i,
  \label{eq:MBLmodel}
\end{align}
where $\boldsymbol{S}$ denote spin-$1/2$ operators. The local fields $h^\alpha_i$ are drawn from a uniform box distribution with zero mean and width $h^\alpha_{\textrm{max}}$. We set $h^x_{\textrm{max}} = h^z_{\textrm{max}} = h_{\textrm{max}}$ and $h^y_{\textrm{max}} = 0$. The disorder allows us to generate many samples at a fixed set of model parameters, in analogy to the different configurations for a fixed temperature in the classical spin systems \cite{Car16a, Wan16a}. 

The model in equation (\ref{eq:MBLmodel}) has a transition between thermalizing and non-thermalizing (i.e. many-body localized) behavior, driven by the disorder strength $h_{\textrm{max}}$. In particular, when varying $h_{\textrm{max}}$, both the energy level statistics as well as the statistics of the entanglement spectra change their nature \cite{Geraedts2016}. For the case of the energy levels, the gaps (level spacings) follow either a Wigner-Dyson distribution for the thermalizing phase,  or a Poisson distribution for the localized phase; while for the entanglement spectrum, the Wigner-Dyson distribution is replaced by a semi-Poisson distribution. Note that the change of ES can already be seen from the statistics in a \emph{single} eigenstate \cite{Geraedts2016}. 

We numerically obtain the entanglement spectrum for the groundstate of the model in equation (\ref{eq:MBLmodel}), for disorder strengths between $h_{\textrm{max}} = J$ and $h_{\textrm{max}} = 5J$. The transition was shown to happen around $h_{\textrm{max}} \approx 3J$ \cite{Geraedts2016}, but we stress that our method does not rely on this knowledge. We would simply have started from a larger width of points, and then systematically narrow it down to the current range. At each value of $h_{\textrm{max}}$ we generate $10^5$ disorder realizations for system size $L = 12$ and calculate the entanglement spectrum for $L_A=L_B=6$. These $2^6=64$ levels are used as the input to the NN. 

First, we try to use an unsupervised PCA to cluster the data. This analysis shows that the first two principal components are dominant, with the other components being of order $10^{-4}$ or less. However, a scatterplot of the data when projected onto the first two principal components (shown in Fig.~\ref{fig:MBLPCA}a) does not reveal a clear clustering of the spectra.

We therefore turn to train a shallow feedforward network on the entanglement spectra in order to use the confusion scheme. Here we use a network with 64 input neurons, 100 hidden neurons and 2 output neurons. The results are shown in Fig.~\ref{fig:MBLPerformance}b. Also in this case, the characteristic W-shape is obtained and we detect the transition at $h_c \approx 3J$. In addition to the previous cases, we also consider explicitly the performance of the network $\mathcal{N}_{h'_{c}}$ \emph{at} $h'_c$. We do this to confirm that the labelling with $h'_c$ at $3J$ is indeed correct. We expect namely that the training of the network is most robust against changes in its parameters for the correct labelling. In other words, we may also look for the $h'_c$ at which the training is most independent of chosen conditions. As shown in Fig.~\ref{fig:MBLPerformance}c, this point is also at $h'_c \approx 3J$.

\section{Discussion}
In this work we have explored the uses of machine learning in identifying (quantum) phase transitions from the overall performance of neural networks. As input data for training we proposed to use the entanglement spectrum of a quantum state, since it provides an excellent way of compressing the otherwise exponentially large wavefunction. We have shown that by confusing the neural network, by purposefully mislabelling the data, we are able to identify the phase transition between two phases of matter without prior knowledge of them. We demonstrated this for three different scenario's. We expect our method to be useful also in situations where it is not clear whether two phases are identical, or whether intervening phases exist in between other phases. If the underlying data in such situations has a (hidden) pattern that distinguishes them, our machine learning approach may well be able to find it. Our method is not limited to applications in condensed matter physics and phase transitions, but might prove useful in any field in which large amounts of data can be classified using a priori unknown labels. 

An interesting direction for future studies is the relaxation of the assumption that there are only two phases to be distinguished. If there are multiple phase transitions present in the data, the characteristic W-shape will be modified, and its new shape (i.e. the number of peaks) will signal the correct number of different labels. Additionally, it may be possible to formulate this method in a self-consistent way, with an adaptive labelling and having the algorithm determine the correct labels by itself. 

\section*{Acknowledgements}
E.v.N and S.H. gratefully acknowledge financial support from the Swiss National Science Foundation (SNSF). Y.-H.L. is supported by ERC Advanced Grant SIMCOFE. E.v.N. acknowledges fruitful discussions with Maciej Koch-Janusz on extending the confusion scheme to the case with multiple phases. E.v.N. and Y.-H.L. acknowledge helpful discussions with Giuseppe Carleo, Juan Osorio and Lei Wang. E.v.N and S.H. thank Andreas Krause for useful discussion on machine learning. \\


\begin{thebibliography}{34}%
\makeatletter
\providecommand \@ifxundefined [1]{%
 \@ifx{#1\undefined}
}%
\providecommand \@ifnum [1]{%
 \ifnum #1\expandafter \@firstoftwo
 \else \expandafter \@secondoftwo
 \fi
}%
\providecommand \@ifx [1]{%
 \ifx #1\expandafter \@firstoftwo
 \else \expandafter \@secondoftwo
 \fi
}%
\providecommand \natexlab [1]{#1}%
\providecommand \enquote  [1]{``#1''}%
\providecommand \bibnamefont  [1]{#1}%
\providecommand \bibfnamefont [1]{#1}%
\providecommand \citenamefont [1]{#1}%
\providecommand \href@noop [0]{\@secondoftwo}%
\providecommand \href [0]{\begingroup \@sanitize@url \@href}%
\providecommand \@href[1]{\@@startlink{#1}\@@href}%
\providecommand \@@href[1]{\endgroup#1\@@endlink}%
\providecommand \@sanitize@url [0]{\catcode `\\12\catcode `\$12\catcode
  `\&12\catcode `\#12\catcode `\^12\catcode `\_12\catcode `\%12\relax}%
\providecommand \@@startlink[1]{}%
\providecommand \@@endlink[0]{}%
\providecommand \url  [0]{\begingroup\@sanitize@url \@url }%
\providecommand \@url [1]{\endgroup\@href {#1}{\urlprefix }}%
\providecommand \urlprefix  [0]{URL }%
\providecommand \Eprint [0]{\href }%
\providecommand \doibase [0]{http://dx.doi.org/}%
\providecommand \selectlanguage [0]{\@gobble}%
\providecommand \bibinfo  [0]{\@secondoftwo}%
\providecommand \bibfield  [0]{\@secondoftwo}%
\providecommand \translation [1]{[#1]}%
\providecommand \BibitemOpen [0]{}%
\providecommand \bibitemStop [0]{}%
\providecommand \bibitemNoStop [0]{.\EOS\space}%
\providecommand \EOS [0]{\spacefactor3000\relax}%
\providecommand \BibitemShut  [1]{\csname bibitem#1\endcsname}%
\let\auto@bib@innerbib\@empty
\bibitem [{\citenamefont {Jordan}\ and\ \citenamefont
  {Mitchell}(2015)}]{Jordan2015}%
  \BibitemOpen
  \bibfield  {author} {\bibinfo {author} {\bibfnamefont {M.~I.}\ \bibnamefont
  {Jordan}}\ and\ \bibinfo {author} {\bibfnamefont {T.~M.}\ \bibnamefont
  {Mitchell}},\ }\href {\doibase 10.1126/science.aaa8415} {\bibfield  {journal}
  {\bibinfo  {journal} {Science}\ }\textbf {\bibinfo {volume} {349}},\ \bibinfo
  {pages} {255} (\bibinfo {year} {2015})}\BibitemShut {NoStop}%
\bibitem [{\citenamefont {Wang}(2016)}]{Wan16a}%
  \BibitemOpen
  \bibfield  {author} {\bibinfo {author} {\bibfnamefont {L.}~\bibnamefont
  {Wang}},\ }\href {https://arxiv.org/abs/1606.00318} {\bibfield  {journal}
  {\bibinfo  {journal} {Preprint at https://arxiv.org/abs/1606.00318}\ }
  (\bibinfo {year} {2016})}\BibitemShut {NoStop}%
\bibitem [{\citenamefont {Carrasquilla}\ and\ \citenamefont
  {Melko}(2016)}]{Car16a}%
  \BibitemOpen
  \bibfield  {author} {\bibinfo {author} {\bibfnamefont {J.}~\bibnamefont
  {Carrasquilla}}\ and\ \bibinfo {author} {\bibfnamefont {R.~G.}\ \bibnamefont
  {Melko}},\ }\href {https://arxiv.org/abs/1605.01735} {\bibfield  {journal}
  {\bibinfo  {journal} {Preprint at https://arxiv.org/abs/1605.01735}\ }
  (\bibinfo {year} {2016})}\BibitemShut {NoStop}%
\bibitem [{\citenamefont {Ch'ng}\ \emph {et~al.}(2016)\citenamefont {Ch'ng},
  \citenamefont {Carrasquilla}, \citenamefont {Melko},\ and\ \citenamefont
  {Khatami}}]{Chng2016}%
  \BibitemOpen
  \bibfield  {author} {\bibinfo {author} {\bibfnamefont {K.}~\bibnamefont
  {Ch'ng}}, \bibinfo {author} {\bibfnamefont {J.}~\bibnamefont {Carrasquilla}},
  \bibinfo {author} {\bibfnamefont {R.~G.}\ \bibnamefont {Melko}}, \ and\
  \bibinfo {author} {\bibfnamefont {E.}~\bibnamefont {Khatami}},\ }\href
  {http://arxiv.org/abs/1609.02552} {\bibfield  {journal} {\bibinfo  {journal}
  {Preprint at http://arxiv.org/abs/1609.02552}\ } (\bibinfo {year}
  {2016})}\BibitemShut {NoStop}%
\bibitem [{\citenamefont {Li}\ \emph {et~al.}(2016)\citenamefont {Li},
  \citenamefont {Baker}, \citenamefont {White},\ and\ \citenamefont
  {Burke}}]{Li2016b}%
  \BibitemOpen
  \bibfield  {author} {\bibinfo {author} {\bibfnamefont {L.}~\bibnamefont
  {Li}}, \bibinfo {author} {\bibfnamefont {T.~E.}\ \bibnamefont {Baker}},
  \bibinfo {author} {\bibfnamefont {S.~R.}\ \bibnamefont {White}}, \ and\
  \bibinfo {author} {\bibfnamefont {K.}~\bibnamefont {Burke}},\ }\href
  {http://arxiv.org/abs/1609.03705} {\bibfield  {journal} {\bibinfo  {journal}
  {Preprint at http://arxiv.org/abs/1609.03705}\ } (\bibinfo {year}
  {2016})}\BibitemShut {NoStop}%
\bibitem [{\citenamefont {Broecker}\ \emph {et~al.}(2016)\citenamefont
  {Broecker}, \citenamefont {Carrasquilla}, \citenamefont {Melko},\ and\
  \citenamefont {Trebst}}]{Broecker2016}%
  \BibitemOpen
  \bibfield  {author} {\bibinfo {author} {\bibfnamefont {P.}~\bibnamefont
  {Broecker}}, \bibinfo {author} {\bibfnamefont {J.}~\bibnamefont
  {Carrasquilla}}, \bibinfo {author} {\bibfnamefont {R.~G.}\ \bibnamefont
  {Melko}}, \ and\ \bibinfo {author} {\bibfnamefont {S.}~\bibnamefont
  {Trebst}},\ }\href {http://arxiv.org/abs/1608.07848} {\bibfield  {journal}
  {\bibinfo  {journal} {Preprint at http://arxiv.org/abs/1608.07848}\ }
  (\bibinfo {year} {2016})}\BibitemShut {NoStop}%
\bibitem [{\citenamefont {Carleo}\ and\ \citenamefont
  {Troyer}(2016)}]{Carleo16a}%
  \BibitemOpen
  \bibfield  {author} {\bibinfo {author} {\bibfnamefont {G.}~\bibnamefont
  {Carleo}}\ and\ \bibinfo {author} {\bibfnamefont {M.}~\bibnamefont
  {Troyer}},\ }\href {https://arxiv.org/abs/1606.02318} {\bibfield  {journal}
  {\bibinfo  {journal} {Preprint at https://arxiv.org/abs/1606.02318}\ }
  (\bibinfo {year} {2016})}\BibitemShut {NoStop}%
\bibitem [{\citenamefont {Deng}\ \emph {et~al.}(2016)\citenamefont {Deng},
  \citenamefont {Li},\ and\ \citenamefont {Sarma}}]{Deng2016}%
  \BibitemOpen
  \bibfield  {author} {\bibinfo {author} {\bibfnamefont {D.-L.}\ \bibnamefont
  {Deng}}, \bibinfo {author} {\bibfnamefont {X.}~\bibnamefont {Li}}, \ and\
  \bibinfo {author} {\bibfnamefont {S.~D.}\ \bibnamefont {Sarma}},\ }\href
  {http://arxiv.org/abs/1609.09060} {\bibfield  {journal} {\bibinfo  {journal}
  {Preprint at http://arxiv.org/abs/1609.09060}\ } (\bibinfo {year}
  {2016})}\BibitemShut {NoStop}%
\bibitem [{\citenamefont {Mehta}\ and\ \citenamefont {Schwab}(2014)}]{Pan14a}%
  \BibitemOpen
  \bibfield  {author} {\bibinfo {author} {\bibfnamefont {P.}~\bibnamefont
  {Mehta}}\ and\ \bibinfo {author} {\bibfnamefont {D.~J.}\ \bibnamefont
  {Schwab}},\ }\href {http://arxiv.org/abs/1410.3831} {\bibfield  {journal}
  {\bibinfo  {journal} {Preprint at http://arxiv.org/abs/1410.3831}\ }
  (\bibinfo {year} {2014})}\BibitemShut {NoStop}%
\bibitem [{\citenamefont {Stoudenmire}\ and\ \citenamefont
  {Schwab}(2016)}]{Sto16a}%
  \BibitemOpen
  \bibfield  {author} {\bibinfo {author} {\bibfnamefont {E.~M.}\ \bibnamefont
  {Stoudenmire}}\ and\ \bibinfo {author} {\bibfnamefont {D.~J.}\ \bibnamefont
  {Schwab}},\ }\href {https://arxiv.org/abs/1605.05775} {\bibfield  {journal}
  {\bibinfo  {journal} {Preprint at https://arxiv.org/abs/1605.05775}\ }
  (\bibinfo {year} {2016})}\BibitemShut {NoStop}%
\bibitem [{\citenamefont {Saitta}\ and\ \citenamefont
  {Sebag}(2010)}]{Saitta2010}%
  \BibitemOpen
  \bibfield  {author} {\bibinfo {author} {\bibfnamefont {L.}~\bibnamefont
  {Saitta}}\ and\ \bibinfo {author} {\bibfnamefont {M.}~\bibnamefont {Sebag}},\
  }\href {\doibase 10.1007/978-0-387-30164-8_635} {\emph {\bibinfo {title}
  {Encyclopedia of Machine Learning}}}\ (\bibinfo  {publisher} {Springer US},\
  \bibinfo {address} {Boston, MA},\ \bibinfo {year} {2010})\ pp.\ \bibinfo
  {pages} {767--773}\BibitemShut {NoStop}%
\bibitem [{\citenamefont {Haykin}(1998)}]{Hay98a}%
  \BibitemOpen
  \bibfield  {author} {\bibinfo {author} {\bibfnamefont {S.~O.}\ \bibnamefont
  {Haykin}},\ }\href@noop {} {\emph {\bibinfo {title} {Neural Networks: A
  Comprehensive Foundation}}}\ (\bibinfo  {publisher} {Prentice Hall},\
  \bibinfo {year} {1998})\BibitemShut {NoStop}%
\bibitem [{\citenamefont {Li}\ and\ \citenamefont {Haldane}(2008)}]{Li2008}%
  \BibitemOpen
  \bibfield  {author} {\bibinfo {author} {\bibfnamefont {H.}~\bibnamefont
  {Li}}\ and\ \bibinfo {author} {\bibfnamefont {F.~D.~M.}\ \bibnamefont
  {Haldane}},\ }\href {\doibase 10.1103/PhysRevLett.101.010504} {\bibfield
  {journal} {\bibinfo  {journal} {Phys. Rev. Lett.}\ }\textbf {\bibinfo
  {volume} {101}},\ \bibinfo {pages} {010504} (\bibinfo {year}
  {2008})}\BibitemShut {NoStop}%
\bibitem [{\citenamefont {Laflorencie}(2015)}]{Laflorencie2015}%
  \BibitemOpen
  \bibfield  {author} {\bibinfo {author} {\bibfnamefont {N.}~\bibnamefont
  {Laflorencie}},\ }\href {http://arxiv.org/abs/1512.03388} {\bibfield
  {journal} {\bibinfo  {journal} {Preprint at
  https://arxiv.org/abs/1512.03388}\ } (\bibinfo {year} {2015})}\BibitemShut
  {NoStop}%
\bibitem [{\citenamefont {Chandran}\ \emph {et~al.}(2014)\citenamefont
  {Chandran}, \citenamefont {Khemani},\ and\ \citenamefont {Sondhi}}]{Cha14a}%
  \BibitemOpen
  \bibfield  {author} {\bibinfo {author} {\bibfnamefont {A.}~\bibnamefont
  {Chandran}}, \bibinfo {author} {\bibfnamefont {V.}~\bibnamefont {Khemani}}, \
  and\ \bibinfo {author} {\bibfnamefont {S.~L.}\ \bibnamefont {Sondhi}},\
  }\href {\doibase 10.1103/PhysRevLett.113.060501} {\bibfield  {journal}
  {\bibinfo  {journal} {Phys. Rev. Lett.}\ }\textbf {\bibinfo {volume} {113}},\
  \bibinfo {pages} {060501} (\bibinfo {year} {2014})}\BibitemShut {NoStop}%
\bibitem [{\citenamefont {Amico}\ \emph {et~al.}(2008)\citenamefont {Amico},
  \citenamefont {Fazio}, \citenamefont {Osterloh},\ and\ \citenamefont
  {Vedral}}]{Amico2008}%
  \BibitemOpen
  \bibfield  {author} {\bibinfo {author} {\bibfnamefont {L.}~\bibnamefont
  {Amico}}, \bibinfo {author} {\bibfnamefont {R.}~\bibnamefont {Fazio}},
  \bibinfo {author} {\bibfnamefont {A.}~\bibnamefont {Osterloh}}, \ and\
  \bibinfo {author} {\bibfnamefont {V.}~\bibnamefont {Vedral}},\ }\href
  {\doibase 10.1103/RevModPhys.80.517} {\bibfield  {journal} {\bibinfo
  {journal} {Rev. Mod. Phys.}\ }\textbf {\bibinfo {volume} {80}},\ \bibinfo
  {pages} {517} (\bibinfo {year} {2008})}\BibitemShut {NoStop}%
\bibitem [{\citenamefont {Thomale}\ \emph
  {et~al.}(2010{\natexlab{a}})\citenamefont {Thomale}, \citenamefont
  {Sterdyniak}, \citenamefont {Regnault},\ and\ \citenamefont
  {Bernevig}}]{Thomale2010}%
  \BibitemOpen
  \bibfield  {author} {\bibinfo {author} {\bibfnamefont {R.}~\bibnamefont
  {Thomale}}, \bibinfo {author} {\bibfnamefont {A.}~\bibnamefont {Sterdyniak}},
  \bibinfo {author} {\bibfnamefont {N.}~\bibnamefont {Regnault}}, \ and\
  \bibinfo {author} {\bibfnamefont {B.~A.}\ \bibnamefont {Bernevig}},\ }\href
  {\doibase 10.1103/PhysRevLett.104.180502} {\bibfield  {journal} {\bibinfo
  {journal} {Phys. Rev. Lett.}\ }\textbf {\bibinfo {volume} {104}},\ \bibinfo
  {pages} {180502} (\bibinfo {year} {2010}{\natexlab{a}})}\BibitemShut
  {NoStop}%
\bibitem [{\citenamefont {Qi}\ \emph {et~al.}(2012)\citenamefont {Qi},
  \citenamefont {Katsura},\ and\ \citenamefont {Ludwig}}]{Qi2012}%
  \BibitemOpen
  \bibfield  {author} {\bibinfo {author} {\bibfnamefont {X.~L.}\ \bibnamefont
  {Qi}}, \bibinfo {author} {\bibfnamefont {H.}~\bibnamefont {Katsura}}, \ and\
  \bibinfo {author} {\bibfnamefont {A.~W.~W.}\ \bibnamefont {Ludwig}},\ }\href
  {\doibase 10.1103/PhysRevLett.108.196402} {\bibfield  {journal} {\bibinfo
  {journal} {Phys. Rev. Lett.}\ }\textbf {\bibinfo {volume} {108}},\ \bibinfo
  {pages} {1} (\bibinfo {year} {2012})}\BibitemShut {NoStop}%
\bibitem [{\citenamefont {Turner}\ \emph {et~al.}(2011)\citenamefont {Turner},
  \citenamefont {Pollmann},\ and\ \citenamefont {Berg}}]{Pol11a}%
  \BibitemOpen
  \bibfield  {author} {\bibinfo {author} {\bibfnamefont {A.~M.}\ \bibnamefont
  {Turner}}, \bibinfo {author} {\bibfnamefont {F.}~\bibnamefont {Pollmann}}, \
  and\ \bibinfo {author} {\bibfnamefont {E.}~\bibnamefont {Berg}},\ }\href
  {\doibase 10.1103/PhysRevB.83.075102} {\bibfield  {journal} {\bibinfo
  {journal} {Phys. Rev. B}\ }\textbf {\bibinfo {volume} {83}},\ \bibinfo
  {pages} {075102} (\bibinfo {year} {2011})}\BibitemShut {NoStop}%
\bibitem [{\citenamefont {Cirac}\ \emph {et~al.}(2011)\citenamefont {Cirac},
  \citenamefont {Poilblanc}, \citenamefont {Schuch},\ and\ \citenamefont
  {Verstraete}}]{Cirac2011}%
  \BibitemOpen
  \bibfield  {author} {\bibinfo {author} {\bibfnamefont {J.~I.}\ \bibnamefont
  {Cirac}}, \bibinfo {author} {\bibfnamefont {D.}~\bibnamefont {Poilblanc}},
  \bibinfo {author} {\bibfnamefont {N.}~\bibnamefont {Schuch}}, \ and\ \bibinfo
  {author} {\bibfnamefont {F.}~\bibnamefont {Verstraete}},\ }\href {\doibase
  10.1103/PhysRevB.83.245134} {\bibfield  {journal} {\bibinfo  {journal} {Phys.
  Rev. B}\ }\textbf {\bibinfo {volume} {83}},\ \bibinfo {pages} {245134}
  (\bibinfo {year} {2011})}\BibitemShut {NoStop}%
\bibitem [{\citenamefont {Schuch}\ \emph {et~al.}(2013)\citenamefont {Schuch},
  \citenamefont {Poilblanc}, \citenamefont {Cirac},\ and\ \citenamefont
  {P{\'{e}}rez-Garc{\'{i}}a}}]{Schuch2013}%
  \BibitemOpen
  \bibfield  {author} {\bibinfo {author} {\bibfnamefont {N.}~\bibnamefont
  {Schuch}}, \bibinfo {author} {\bibfnamefont {D.}~\bibnamefont {Poilblanc}},
  \bibinfo {author} {\bibfnamefont {J.~I.}\ \bibnamefont {Cirac}}, \ and\
  \bibinfo {author} {\bibfnamefont {D.}~\bibnamefont
  {P{\'{e}}rez-Garc{\'{i}}a}},\ }\href {\doibase
  10.1103/PhysRevLett.111.090501} {\bibfield  {journal} {\bibinfo  {journal}
  {Phys. Rev. Lett.}\ }\textbf {\bibinfo {volume} {111}},\ \bibinfo {pages}
  {090501} (\bibinfo {year} {2013})}\BibitemShut {NoStop}%
\bibitem [{\citenamefont {Calabrese}\ and\ \citenamefont
  {Lefevre}(2008)}]{Calabrese2008a}%
  \BibitemOpen
  \bibfield  {author} {\bibinfo {author} {\bibfnamefont {P.}~\bibnamefont
  {Calabrese}}\ and\ \bibinfo {author} {\bibfnamefont {A.}~\bibnamefont
  {Lefevre}},\ }\href {\doibase 10.1103/PhysRevA.78.032329} {\bibfield
  {journal} {\bibinfo  {journal} {Phys. Rev. A}\ }\textbf {\bibinfo {volume}
  {78}},\ \bibinfo {pages} {032329} (\bibinfo {year} {2008})}\BibitemShut
  {NoStop}%
\bibitem [{\citenamefont {Alba}\ \emph {et~al.}(2012)\citenamefont {Alba},
  \citenamefont {Haque},\ and\ \citenamefont {L\"auchli}}]{Alba2012a}%
  \BibitemOpen
  \bibfield  {author} {\bibinfo {author} {\bibfnamefont {V.}~\bibnamefont
  {Alba}}, \bibinfo {author} {\bibfnamefont {M.}~\bibnamefont {Haque}}, \ and\
  \bibinfo {author} {\bibfnamefont {A.~M.}\ \bibnamefont {L\"auchli}},\ }\href
  {\doibase 10.1103/PhysRevLett.108.227201} {\bibfield  {journal} {\bibinfo
  {journal} {Phys. Rev. Lett.}\ }\textbf {\bibinfo {volume} {108}},\ \bibinfo
  {pages} {227201} (\bibinfo {year} {2012})}\BibitemShut {NoStop}%
\bibitem [{\citenamefont {Yang}\ \emph {et~al.}(2015)\citenamefont {Yang},
  \citenamefont {Chamon}, \citenamefont {Hamma},\ and\ \citenamefont
  {Mucciolo}}]{Yan15a}%
  \BibitemOpen
  \bibfield  {author} {\bibinfo {author} {\bibfnamefont {Z.-C.}\ \bibnamefont
  {Yang}}, \bibinfo {author} {\bibfnamefont {C.}~\bibnamefont {Chamon}},
  \bibinfo {author} {\bibfnamefont {A.}~\bibnamefont {Hamma}}, \ and\ \bibinfo
  {author} {\bibfnamefont {E.~R.}\ \bibnamefont {Mucciolo}},\ }\href {\doibase
  10.1103/PhysRevLett.115.267206} {\bibfield  {journal} {\bibinfo  {journal}
  {Phys. Rev. Lett.}\ }\textbf {\bibinfo {volume} {115}},\ \bibinfo {pages}
  {267206} (\bibinfo {year} {2015})}\BibitemShut {NoStop}%
\bibitem [{\citenamefont {Geraedts}\ \emph {et~al.}(2016)\citenamefont
  {Geraedts}, \citenamefont {Nandkishore},\ and\ \citenamefont
  {Regnault}}]{Geraedts2016}%
  \BibitemOpen
  \bibfield  {author} {\bibinfo {author} {\bibfnamefont {S.~D.}\ \bibnamefont
  {Geraedts}}, \bibinfo {author} {\bibfnamefont {R.}~\bibnamefont
  {Nandkishore}}, \ and\ \bibinfo {author} {\bibfnamefont {N.}~\bibnamefont
  {Regnault}},\ }\href {\doibase 10.1103/PhysRevB.93.174202} {\bibfield
  {journal} {\bibinfo  {journal} {Phys. Rev. B}\ }\textbf {\bibinfo {volume}
  {93}},\ \bibinfo {pages} {174202} (\bibinfo {year} {2016})}\BibitemShut
  {NoStop}%
\bibitem [{\citenamefont {Pichler}\ \emph {et~al.}(2016)\citenamefont
  {Pichler}, \citenamefont {Zhu}, \citenamefont {Seif}, \citenamefont
  {Zoller},\ and\ \citenamefont {Hafezi}}]{Pic16a}%
  \BibitemOpen
  \bibfield  {author} {\bibinfo {author} {\bibfnamefont {H.}~\bibnamefont
  {Pichler}}, \bibinfo {author} {\bibfnamefont {G.}~\bibnamefont {Zhu}},
  \bibinfo {author} {\bibfnamefont {A.}~\bibnamefont {Seif}}, \bibinfo {author}
  {\bibfnamefont {P.}~\bibnamefont {Zoller}}, \ and\ \bibinfo {author}
  {\bibfnamefont {M.}~\bibnamefont {Hafezi}},\ }\href
  {http://arxiv.org/abs/1605.08624} {\bibfield  {journal} {\bibinfo  {journal}
  {Preprint at http://arxiv.org/abs/1605.08624}\ } (\bibinfo {year}
  {2016})}\BibitemShut {NoStop}%
\bibitem [{\citenamefont {Kitaev}(2001)}]{Kitaev2001}%
  \BibitemOpen
  \bibfield  {author} {\bibinfo {author} {\bibfnamefont {A.~Y.}\ \bibnamefont
  {Kitaev}},\ }\href@noop {} {\bibfield  {journal} {\bibinfo  {journal}
  {Physics-Uspekhi}\ }\textbf {\bibinfo {volume} {44}},\ \bibinfo {pages} {131}
  (\bibinfo {year} {2001})}\BibitemShut {NoStop}%
\bibitem [{\citenamefont {Hsieh}\ and\ \citenamefont {Fu}(2014)}]{Hsieh14}%
  \BibitemOpen
  \bibfield  {author} {\bibinfo {author} {\bibfnamefont {T.~H.}\ \bibnamefont
  {Hsieh}}\ and\ \bibinfo {author} {\bibfnamefont {L.}~\bibnamefont {Fu}},\
  }\href {\doibase 10.1103/PhysRevLett.113.106801} {\bibfield  {journal}
  {\bibinfo  {journal} {Phys. Rev. Lett.}\ }\textbf {\bibinfo {volume} {113}},\
  \bibinfo {pages} {106801} (\bibinfo {year} {2014})}\BibitemShut {NoStop}%
\bibitem [{\citenamefont {Thomale}\ \emph
  {et~al.}(2010{\natexlab{b}})\citenamefont {Thomale}, \citenamefont {Arovas},\
  and\ \citenamefont {Bernevig}}]{Thomale2010a}%
  \BibitemOpen
  \bibfield  {author} {\bibinfo {author} {\bibfnamefont {R.}~\bibnamefont
  {Thomale}}, \bibinfo {author} {\bibfnamefont {D.~P.}\ \bibnamefont {Arovas}},
  \ and\ \bibinfo {author} {\bibfnamefont {B.~A.}\ \bibnamefont {Bernevig}},\
  }\href {\doibase 10.1103/PhysRevLett.105.116805} {\bibfield  {journal}
  {\bibinfo  {journal} {Phys. Rev. Lett.}\ }\textbf {\bibinfo {volume} {105}},\
  \bibinfo {pages} {116805} (\bibinfo {year} {2010}{\natexlab{b}})}\BibitemShut
  {NoStop}%
\bibitem [{\citenamefont {Vijay}\ and\ \citenamefont {Fu}(2015)}]{Vijay15}%
  \BibitemOpen
  \bibfield  {author} {\bibinfo {author} {\bibfnamefont {S.}~\bibnamefont
  {Vijay}}\ and\ \bibinfo {author} {\bibfnamefont {L.}~\bibnamefont {Fu}},\
  }\href {\doibase 10.1103/PhysRevB.91.220101} {\bibfield  {journal} {\bibinfo
  {journal} {Phys. Rev. B}\ }\textbf {\bibinfo {volume} {91}},\ \bibinfo
  {pages} {220101} (\bibinfo {year} {2015})}\BibitemShut {NoStop}%
\bibitem [{\citenamefont {Pearson}(1901)}]{Pearson1901}%
  \BibitemOpen
  \bibfield  {author} {\bibinfo {author} {\bibfnamefont {F.~K.}\ \bibnamefont
  {Pearson}},\ }\href {\doibase 10.1080/14786440109462720} {\bibfield
  {journal} {\bibinfo  {journal} {Philosophical Magazine Series 6}\ }\textbf
  {\bibinfo {volume} {2}},\ \bibinfo {pages} {559} (\bibinfo {year}
  {1901})}\BibitemShut {NoStop}%
\bibitem [{\citenamefont {Nielsen}(2015)}]{Nielsen2015}%
  \BibitemOpen
  \bibfield  {author} {\bibinfo {author} {\bibfnamefont {M.}~\bibnamefont
  {Nielsen}},\ }\href@noop {} {\emph {\bibinfo {title} {Neural Networks and
  Deep Learning}}}\ (\bibinfo  {publisher} {Determination Press},\ \bibinfo
  {year} {2015})\BibitemShut {NoStop}%
\bibitem [{\citenamefont {Onsager}(1944)}]{Onsager1944}%
  \BibitemOpen
  \bibfield  {author} {\bibinfo {author} {\bibfnamefont {L.}~\bibnamefont
  {Onsager}},\ }\href {\doibase 10.1103/PhysRev.65.117} {\bibfield  {journal}
  {\bibinfo  {journal} {Phys. Rev.}\ }\textbf {\bibinfo {volume} {65}},\
  \bibinfo {pages} {117} (\bibinfo {year} {1944})}\BibitemShut {NoStop}%
\bibitem [{\citenamefont {Nandkishore}\ and\ \citenamefont
  {Huse}(2015)}]{Nandkishore2015}%
  \BibitemOpen
  \bibfield  {author} {\bibinfo {author} {\bibfnamefont {R.}~\bibnamefont
  {Nandkishore}}\ and\ \bibinfo {author} {\bibfnamefont {D.~A.}\ \bibnamefont
  {Huse}},\ }\href {\doibase 10.1146/annurev-conmatphys-031214-014726}
  {\bibfield  {journal} {\bibinfo  {journal} {Annual Review of Condensed Matter
  Physics}\ }\textbf {\bibinfo {volume} {6}},\ \bibinfo {pages} {15} (\bibinfo
  {year} {2015})}\BibitemShut {NoStop}%
\end{thebibliography}
%

\end{document}